# Sub-bandgap activated charges transfer in a graphene-MoS$_2$-graphene heterostructure

Sunil Kumar,* Arvind Singh, Anand Nivedan, Sandeep Kumar, Seok Joon Yun, Young Hee Lee, Marc Tondusson, Jérôme Degert, Jean Oberle, and Eric Freysz*

[1]*Femtosecond Spectroscopy and Nonlinear Photonics Laboratory,
Department of Physics, Indian Institute of Technology Delhi, New Delhi 110016, India*
[2]*Department of Energy Science, Sungkyunkwan University (SKKU), Suwon 16419, Republic of Korea*
[3]*Univ. Bordeaux, CNRS, LOMA UMR 5798, 33405 Talence, France*
*kumarsunil@physics.iitd.ac.in; *eric.freysz@u-bordeaux.fr

Monolayers of transition metal dichalcogenides are semiconducting materials which offer many prospects in optoelectronics. A monolayer of molybdenum disulfide (MoS$_2$) has a direct bandgap of 1.88 eV. Hence, when excited with optical photon energies below its bandgap, no photocarriers are generated and a monolayer of MoS$_2$ is not of much use in either photovoltaics or photodetection. Here, we demonstrate that large size MoS$_2$ monolayer sandwiched between two graphene layers makes this heterostructure optically active well below the band gap of MoS$_2$. An ultrafast optical pump-THz probe experiment reveals in real-time, transfer of carriers between graphene and MoS$_2$ monolayer upon photoexcitation with photon energies down to 0.5 eV. It also helps to unravel an unprecedented enhancement in the broadband transient THz response of this tri-layer material system. We propose possible mechanism which can account for this phenomenon. Such specially designed heterostructures, which can be easily built around different transition metal dichalcogenide monolayers, will considerably broaden the scope for modern optoelectronic applications at THz bandwidth.

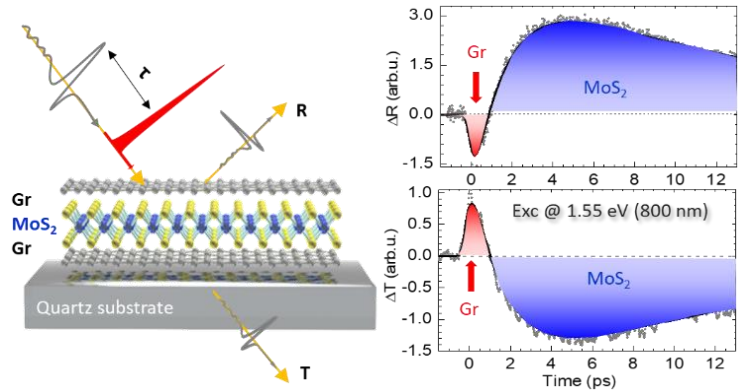

## 1. Introduction

Suitability of optical materials for applications in photovoltaics and photodetection relies on the efficient conversion of light photons into free electron-hole pairs. In this regard the limiting processes related to carrier-carrier scattering, heat generation through phonon emission by hot carriers and the electron-hole recombination, are important to be known as precisely as possible. In real devices, the latter two processes must be reduced to enhance the conversion of light energy into free carriers only. Moreover, to avoid their recombination, the photogenerated carriers need to be spatially separated almost as soon as they are generated. In practice, multi-layers of different materials are often employed for many different purposes. Hence, interfacial effects taking place at ultrafast time-scale and within atomistic regions between the layers, are prone to affect their performances in a crucial way. Recently, novel promising optoelectronic devices using quantum materials such as graphene and transition metal dichalcogenides (TMDs), and their heterostructures have been designed. Heterostructures are of special interest as they serve in the study of the atomic interfacial effects and impact the fundamental time-scales involved during the charge carriers exchange and recombination. In many cases, it also has been observed that optical effects are enhanced by appropriate heterojunction between two different monolayers[1,2] and it happens to be very useful for lot-of applications. In fact, near-field coupling through the interlayer charge and energy transfer processes in metal-semiconductor stacks of two-dimensional materials govern the performance of optoelectronic devices made out of them.[3] Very recently, the attention of the scientific community has been driven towards large area heterostructures of graphene with other monolayers of finite band-gap semiconducting TMDs for novel device applications. The latter are easy to synthesize nowadays and exhibit interesting prospects.[4],[5],[6],[7],[8],[9],[10] Such heterostructures of varying electronic properties allow us to engineer the interactions/coupling between the layers to address specific applications. Thanks to the remarkable optical properties of molybdenum disulfide ($MoS_2$), the high transparency and tunability of the Fermi level in graphene, $MoS_2$-graphene heterostructures have shown good promise for flexible and transparent electronic and optoelectronic applications such as ultrathin vertical transistors,[11,12] field-effect transistors,[13,14] photodetectors,[15] energy convertors,[16] storage devices,[17] ultrafast saturable absorbers[18] and so on[19]. For all these applications, it is important to master the mechanisms involved in the generation of carriers and their dynamics in such heterostructures.

The interaction between graphene and $MoS_2$ monolayer has been studied by different means. Theoretical calculations using density functional theory have suggested that the linear bands at the Dirac points of graphene are slightly changed in the graphene-$MoS_2$ heterostructure.[20–24] In the presence of a single graphene layer, the C-exciton binding energy of $MoS_2$ was reported to be lowered by ~0.45 eV.[25] Transfer of photocarriers from $MoS_2$ to graphene has been reported in the $MoS_2$-graphene heterostructure through transient absorption spectroscopy by exciting the excitonic states of $MoS_2$.[26] Photoluminescence along with Raman spectroscopy have also been used to demonstrate the saturation of charge transfer from $MoSe_2$ to graphene in their heterostructure.[3] In addition to photoluminescence quenching, the life time of the excitons was found to reduce to just a picosecond.[3] However, most of the above studies were performed at optical photon energies either above or near the bandgap of $MoS_2$ and other TMD monolayers. Besides, the techniques involved in these experiments do not directly measure the generated carriers and their dynamics within such heterostructures. In addition, explorations on either TMD "encapsulated" graphene or graphene "encapsulated" TMD monolayer can rarely be found in the literature. Such heterostructures are of particular interest, since a $MoS_2$ monolayer sandwiched between two graphene layers can exhibit dramatic improvement in its stability against defects formation by powerful incident radiation[27]. Recently, a theoretical study on $MoS_2$-graphene-$WX_2$ (X = S and Se) heterostructure showed not only a much higher thermal stability but also an enhanced carrier concentration and electronic transport in this unique tri-layer heterostructure[22]. Similarly, control on the flow of charges, driven by an external field applied perpendicular to graphene-$MoS_2$-graphene and graphene-$MoSe_2$-graphene heterostructures, was also studied theoretically[23,28] to explain the extraordinary performance of such heterostructures for high speed and high on-off switching transistor.[29]

Here, we perform optical pump-THz probe experiments to unravel an extraordinary enhancement in the transient THz response from $MoS_2$ monolayer sandwiched between two graphene layers at optical photon energies in the deep sub-bandgap region of the $MoS_2$ monolayer. The photoinduced dynamic THz conductivity that can be inferred from the transient transmission or reflectivity of the graphene-$MoS_2$-graphene heterostructure, is enhanced significantly and it lasts for a few tens of ps. The graphene-like and $MoS_2$-like contributions can be distinguished easily in the ultrafast response of the heterostructure. We also find that the $MoS_2$-like contribution in the transient THz conductivity of the heterostructure systematically reduces with the decreasing optical excitation photon energy and it becomes negligible at ~0.45 eV. On the other hand, due to negligible THz conductivity of $MoS_2$ monolayer, the stationary THz conductivity of the heterostructure resembles that of graphene in the entire experimental THz bandwidth of ~0.1-5.0 THz.

## 2. Experimental results

### 2.1. THz conductivity measurements

High-quality CVD grown graphene (Gr), $MoS_2$ single layer ($MoS_2SL$) and graphene/$MoS_2SL$/graphene ($GrMoS_2Gr$) were transferred on Quartz substrates. The broad-band static THz sheet conductivity, σ in mS (milli-Siemens) as determined from THz time-domain spectroscopy is presented in FIGURE 1 for these three samples. The procedure is detailed in the supporting information (S4). The thick continuous curves, obtained by polynomial fitting of the data, represent the mean behavior of the spectral conductivity in each case. Re(σ) is the real part and Im(σ) is the imaginary part of the conductivity.

Without a surprise, the THz conductivity of the MoS$_2$ monolayer is almost zero within our experimental uncertainty. This confirms that there are no free carriers present in this high-quality semiconducting monolayer MoS$_2$ sample. For the THz conductivity of the graphene, the results show positive values of the real and imaginary parts in the entire frequency range of ~0.3-5 THz. Likewise, for the GrMoS$_2$Gr, the Re($\sigma$) is almost two times higher than that of graphene, while it continuously decreases with the increasing frequency. For this sample, Im($\sigma$) is negative from ~0.3 THz to 1.7 THz, while it is positive and increasing with the frequency beyond 1.6 THz. The interlayer van der Walls (vdW) interaction in the GrMoS$_2$Gr could be a reason for a smaller value of the Im($\sigma$) as compared to that of graphene and the crossover from small negative values to increasing positive values at a frequency above 1.7 THz.

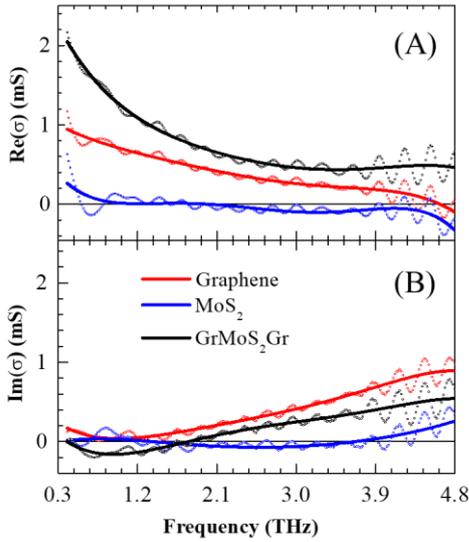

**FIGURE 1** A, Real part, Re($\sigma$) and B, imaginary part, Im($\sigma$) of the broadband static THz sheet conductivity of graphene (Gr), MoS$_2$ monolayer, and GrMoS$_2$Gr heterostructure. Thick continuous curves represent the mean behavior obtained by using polynomial fitting.

### 2.2. Optical pump-induced ultrafast THz response of graphene and MoS$_2$ monolayer

At two optical excitation photon energies (wavelengths) of ~1.55 eV (800 nm) and 3.1 eV (400 nm), the representative optical pump-induced THz reflectivity changes ($\Delta$R) of Gr and MoS$_2$ monolayer, are presented in FIGURES 2A and 2B, respectively, for a nominal optical-fluence of ~20 µJ/cm$^2$. The underlying quartz substrate is completely inactive at both the excitation wavelengths and pump-fluences up to much >500 µJ/cm$^2$. For the graphene, a characteristic negative $\Delta$R response followed by its nearly complete recovery within ~8 ps is observed at both the 1.55 eV and 3.1 eV excitations. Detailed experiments (see FIGURE S6 in the supporting information) have revealed that the $\Delta$R THz response of graphene has mono-exponential recovery with an average time-constant of ~2.8 ps. Fluence-dependent experiments show that the time-constant increases from ~1.5 ps to ~3.5 ps by increasing the fluence from very low values to as high as 300 µJ/cm$^2$. At any given fluence, the dynamics for 3.1 eV is found to be faster than for 1.55 eV, while the saturation of the THz response takes place at a similar fluence of ~60 µJ/cm$^2$. The relaxation of energetic carriers in graphene is known to occur via production of additional electron–hole pairs by carrier–carrier scattering processes rather than heat production by electron-phonon scattering[30].

For the MoS$_2$SL, due to absence of photocarrier generation, there is no photoinduced dynamic THz response at 1.55 eV. Only 3.1 eV excitation produces characteristic positive $\Delta$R whose complete recovery seems to take place via two relaxation channels, a fast one at shorter times and another slower one at the longer times. Both the fast relaxation time-constant (mean value $\tau_1$ ~ 1 ps) and slow relaxation time-constant (mean value $\tau_2$ ~12 ps) show a very weak fluence-dependence in the experimental range of 0-250 µJ cm$^{-2}$ (see FIGURE S7 in the supporting information). The fast relaxation component shows saturation at fluences above ~150 µJ cm$^{-2}$ while the slow component gets saturated at ~50 µJ cm$^{-2}$. It may be pointed out that in comparison to our single fast relaxation component ($\tau_1$~1 ps), two time-constants with values ~350 fs and ~ 1 ps were reported in the literature for MoS$_2$,[31,32] where it was shown that surface trapping is the dominant mechanism of ultrafast THz conductivity quenching.[31] The slower time constant $\tau_2$~12 ps in the present case is found to be about twice faster than that was attributed to trion relaxation earlier.[32] Nevertheless, the dynamics in MoS$_2$ monolayer is found to be similar to that in the MoSe$_2$ monolayer at 3.1 eV photoexcitation.[2]

### 2.3. Optical pump-induced THz response of GrMoS$_2$Gr

The evolution of the GrMoS$_2$Gr reflectivity in the THz range upon excitation at 3.1 eV, i.e., above the bandgap of MoS$_2$, for the pump-fluence of ~14 µJ cm$^{-2}$ is displayed in FIGURE 3A. Upon excitation, the transient response first decreases slightly and then increases rapidly to large values, following which, a slow decrease occurs. An enhancement in the ultrafast THz response at 3.1 eV as compared to that at 1.55 eV, is expected here, because photocarriers are generated in both the $MoS_2$ and graphene layers for the prior. However, the signal cannot naively be accounted for by adding the contributions of graphene and $MoS_2$, displayed in FIGURE 2. Indeed, it has been shown that, upon excitation above its bandgap, the electrons photogenerated in $MoS_2$ are rapidly transferred towards graphene. Fluence-dependent experiments at 3.1 eV reveal that at the zero delay, there appears a contribution from a THz pulse-like response (FIGURE S9 in the supplementary information) that is superimposed with the main THz transient signal of $GrMoS_2Gr$. This additional contribution may arise due to THz emission from the optical pump-induced charge separation at the interface which has been seen earlier in $MoS_2/MoSe_2$ heterostructure[1].

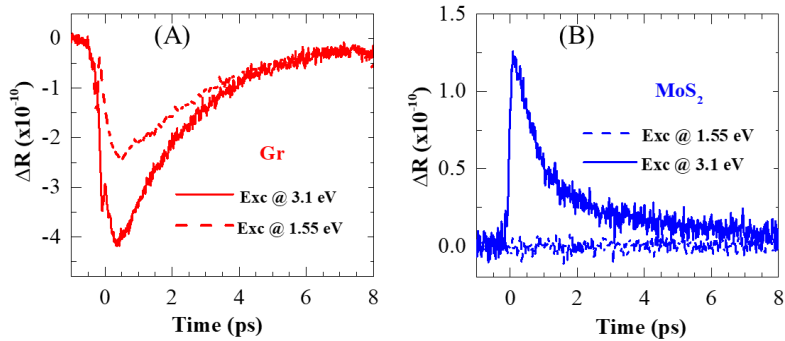

**FIGURE 2** Time-resolved THz reflectivity change, $\Delta$R measured in the lock-in units for A, graphene (Gr), and B, MoS$_2$ monolayer upon femtosecond optical excitation at 3.1 eV and 1.55 eV using pump-fluence of ~20 µJ/cm$^2$.

Typical variations and kinetics of the optical pump-induced THz reflectivity, $\Delta R$ and transmission, $\Delta T$ changes for $GrMoS_2Gr$ are presented in FIGURES 3B and 3C, respectively, at the excitation photon energy of 1.55 eV (below the $MoS_2SL$ bandgap) and fixed fluence of ~20 µJ cm$^{-2}$. It is obvious that, except the polarity reversal, the THz response of the heterostructure through both the $\Delta R$ and $\Delta T$, is exactly the same. Therefore, analysis of any of the two is sufficient for extracting the relaxation kinetics. The characteristic THz responses of both, the graphene and the $MoS_2$ constituents of this tri-layer stack, are observed here and marked by red (Gr-like at short times) and blue ($MoS_2$-like at longer times) in the figure. Compared to the data in FIGURE 2 for the $MoS_2$ monolayer, a surprising enhancement in the $MoS_2$-like contribution in the transient THz reflectivity of the photoexcited $GrMoS_2Gr$ heterostructure, is recorded. This is true for both the excitation photon energies here. Besides, since at 1.55 eV, no photocarriers are generated in $MoS_2SL$, a $MoS_2$-like contribution is not expected in the THz response of $GrMoS_2Gr$ in the first place.

**2.4. Kinetics of the THz response of GrMoS$_2$Gr**

FIGURE 4A displays the temporal evolution of the THz reflectivity, $\Delta R(t)$ of the $GrMoS_2Gr$ sample upon excitation at 1.55 eV taken at various fluences. Similar results were obtained in transmission mode (see supporting information FIGURE S8). The data are well fitted using bi-exponential relaxation components in the convolution. As we already mentioned, the amplitude, A and the relaxation constant, $\tau$ of these components will be considered as Gr-like ($-A_1, \tau_1$) and $MoS_2$-like ($A_2, \tau_2$), respectively. The amplitude $A_1$ of the Gr-like component is negative and it relaxes with a short time constant (1.6 ps<$\tau_1$<1.9 ps). The amplitude $A_2$ of the $MoS_2$-like component is positive and it relaxes with a longer time constant (13 ps<$\tau_2$<16 ps). The fluence-dependence of these parameters is given in FIGURE 4B for the negative Gr-like and positive $MoS_2$-like relaxation components. Like in graphene alone, the Gr-like component in $GrMoS_2Gr$, quickly saturates above a fluence of ~60 µJ/cm$^2$ and the time-constant, $\tau_1$ also shows an increase with the increasing fluence to attain a constant value at very high fluences. The latter is slightly smaller (mean value ~1.8 ps) than the mean value obtained for graphene alone (~2.5 ps). The $MoS_2$-like component is rather interesting. The amplitude $A_2$ shows saturation at fluences above ~150 µJ/cm$^2$, exactly like the behavior of the $MoS_2$ monolayer alone as discussed above for 3.1 eV excitation (also FIGURE S7 of the supplementary information). The corresponding time-constant, $\tau_2$, increases from ~13 ps at the low fluences (same value as in $MoS_2$ alone, FIGURE S7) to ~16 ps at higher fluences in FIGURE 4B. Since the contribution (area under the curve) of the $MoS_2$-like component in the THz response of the $GrMoS_2Gr$, is much larger, a higher value of $\tau_2$ means that the overall THz response from $GrMoS_2Gr$ has got enhanced as compared to that of $MoS_2$ monolayer at 3.1 eV.

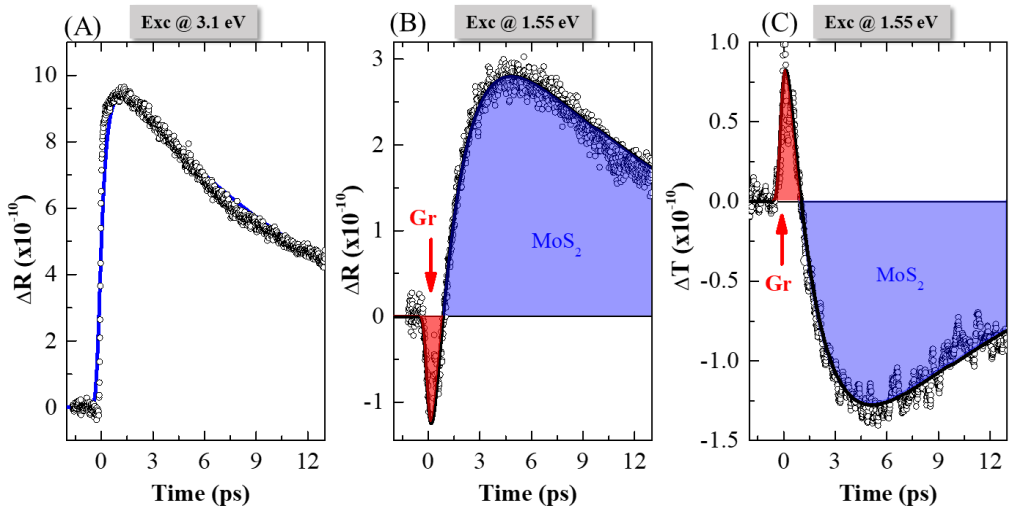

**FIGURE 3** Time-resolved THz response of GrMoS$_2$Gr measured in the lock-in units for A, femtosecond optical excitation at 3.1 eV with a pump-fluence of ~14 µJ/cm$^2$ and measured in the reflection mode. The transient THz response in B, reflection, and C, transmission modes for excitation at 1.55 eV using pump-fluence of ~20 µJ/cm$^2$ revealing Gr-like and MoS$_2$-like contributions.

The transient THz reflectivity data for the GrMoS$_2$Gr at 3.1 eV and a fixed fluence was shown in FIGURE 3A. From the raw traces taken at various fluences (FIGURE S9 in the supplementary information), the Gr-like contribution, just near the zero-delay, appears less prominent due to an overlapping THz pulse-like feature. However, the data beyond time ~0.8 ps can be reproduced well by using a bi-exponential function in the convolution, exactly like what was done for the response at 1.55 eV. The fluence-dependence of the amplitudes and the time-constants of the negative (-A$_1$,τ$_1$) Gr-like and positive (A$_2$,τ$_2$) MoS$_2$-like components in ΔR response at 3.1 eV, is given in FIGURE 4C. As the amplitudes of the two contributions saturate above ~70 μJ/cm$^2$, the time-constant, τ$_1$ of Gr-like component remains almost steady at ~1 ps, while, the time-constant, τ$_2$ of the MoS$_2$-like component sligthly decreases from ~14 ps to ~12 ps. Importantly, at low fluences for both the pump photon energies, 1.55 eV and 3.1 eV, the two time-constants are similar to each other suggesting the same origin. Hence, in absence of any direct photoexcitation of the carriers in the MoS$_2$ layer of the GrMoS$_2$Gr at 1.55 eV, our observations indicate that a significant fraction of the photocarriers generated in the graphene layers migrate towards the MoS$_2$ layer so as to enhance the THz response from GrMoS$_2$Gr. The carriers' transfer to MoS$_2$ is likely to take place during the excitation pulse itself. Later, these carriers will have to return back to graphene for the recombination to take place, thereby slowing down the overall carrier relaxation process in GrMoS$_2$Gr.

### 2.5. Deep sub-bandgap optically-induced THz response of GrMoS$_2$Gr

To strengthen the above observation regarding the THz response of GrMoS$_2$Gr excited below the MoS$_2$-bandgap and get more details before discussing the possible mechanism for this effect, we present here additional experimental results obtained at excitation photon energies much smaller than 1.55 eV. Transient THz transmission response, ΔT of the GrMoS$_2$Gr with excitation photon energies of ~1.1 eV (1140 nm), 0.92 eV (1240 nm) and 0.46 eV (2700 nm), i.e., going deeper into the sub-bandgap region of the MoS$_2$SL are presented in FIGURE 5 taken at a pump fluence of ~50 μJ/cm$^2$. For a one-to-one comparison, the corresponding results for graphene alone, are also presented in FIGURE 5.

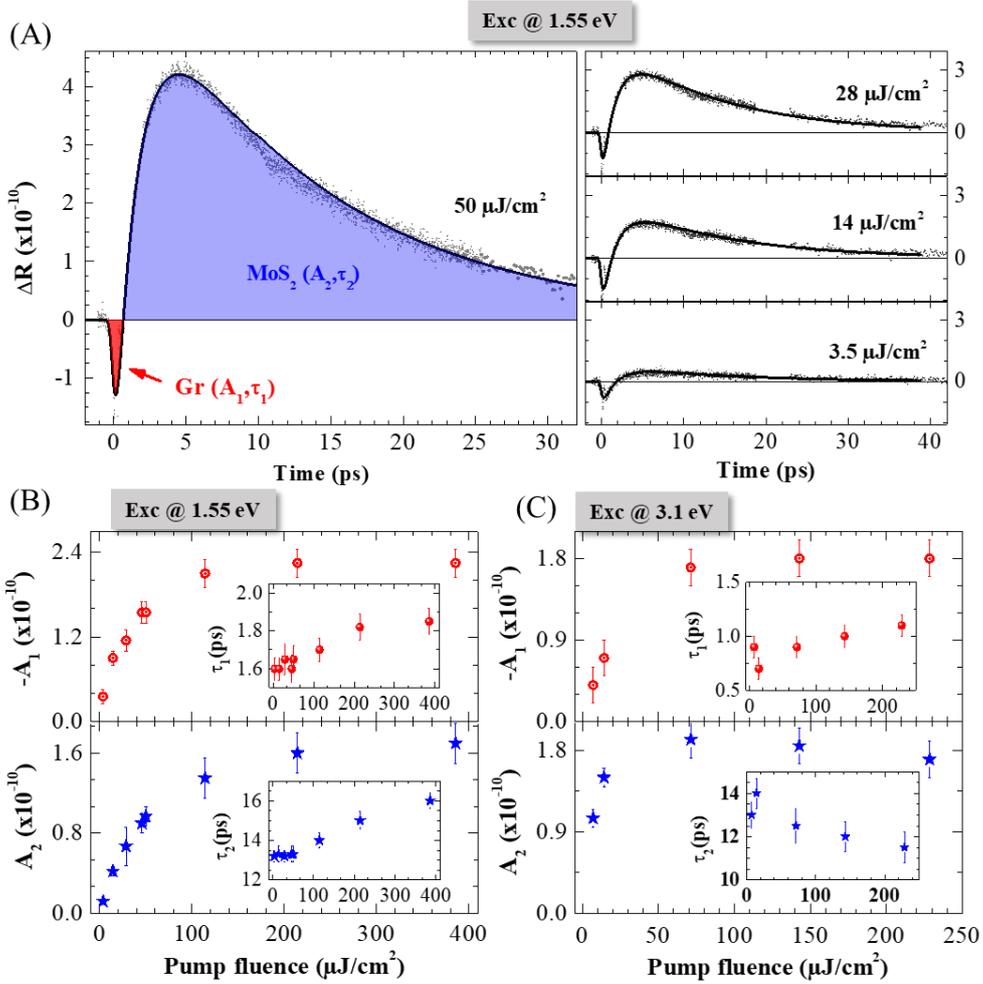

**FIGURE 4** Femtosecond optical pulse-induced modifications in the THz reflectivity, ΔR of GrMoS$_2$Gr heterostructure in the lock-in units. A, Evolution of ΔR dynamics at various pump-fluences following optical excitation at 1.55 eV. The Gr-like and MoS$_2$-like components are marked. The fit of the data and corresponding relaxation components (A$_1$,τ$_1$, A$_2$,τ$_2$) are also indicated. Pump-fluence dependent evolution of the amplitudes and time-constants of the Gr-like (A$_1$,τ$_1$) and MoS$_2$-like (A$_2$,τ$_2$) components for optical excitation at B, 1.55 eV and C, 3.1 eV.

Clearly, the characteristic positive ΔT response of graphene remains intact in GrMoS$_2$Gr at all the excitation wavelengths in FIGURE 5. However, the MoS$_2$-like contribution keeps reducing with the lowering of the excitation photon energy. At excitation photon energy of 0.46 eV, the MoS$_2$-like contribution becomes negligibly small, in comparison, and the overall THz response of GrMoS$_2$Gr quite resembles that of the graphene. It is evident that the MoS$_2$-like contribution will vanish completely, if the excitation photon energy is reduced further down. We may also note from FIGURE 5 that the THz response of graphene at 0.46 eV (2700 nm) possesses a second slow relaxation component with its time-constant of ~10 ps.

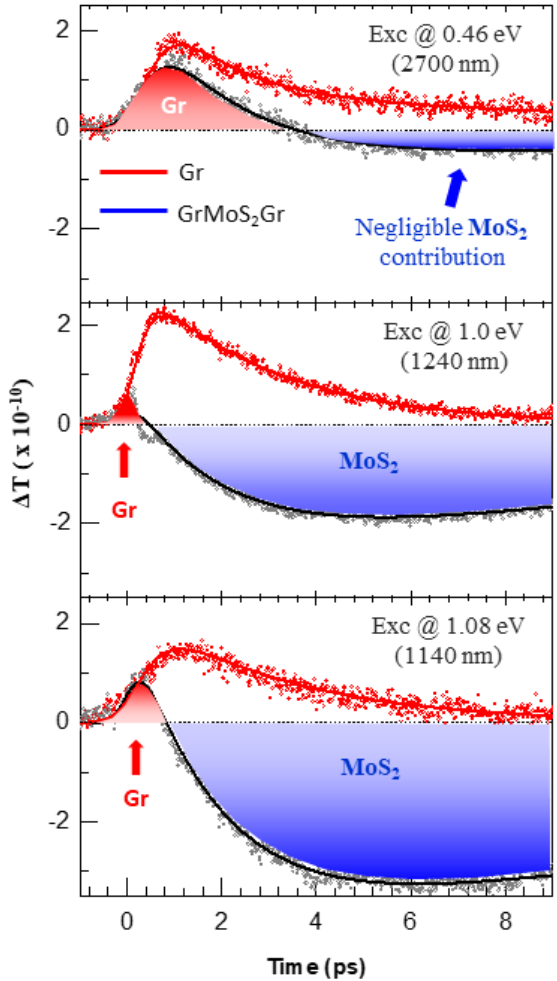

**FIGURE 5** Transient THz transmission changes at three pump photon energies (wavelenths) as marked. For a direct comparision, results for the graphene sample are also shown by the red point (data) and continuous curves (fitting).

### 3. Discussion

A possible mechanism accounting for our experimental observations is provided in FIGURE 6 through charge creation and relaxation in the real and momentum space representations of the GrMoS$_2$Gr. The upper panel in FIGURE 6 displays the physical stack which undergoes optical pulse excitation and THz pulse probing. The free electrons generation in graphene and their back and forth movement between Graphene and MoS$_2$ is indicated. Time-dependent THz absorption by these free carriers is what is captured through the THz probing with controllable delay after the pump pulse excitation. The electronic energy structures of graphene and MoS$_2$ layer along with their density of states (DOS)[33] are drawn in the lower panel. At room temperature, even without any intentional doping, there are a finite number of free carriers in graphene. This could be the reason for the occurrence of the slow relaxation component in the transient THz response of graphene due to optical excitation at 0.46 eV (2700 nm) as discussed in the previous section. In the GrMoS$_2$Gr, the layers are physically separated by about 0.5 nm and electrostatically bonded through vdW interaction. As soon as the interface between graphene and MoS$_2$ layer is formed, the bottom of the conduction band of MoS$_2$ equilibrates at an energy ~0.5 eV above the K-point of graphene[21,23,24] as indicated in the FIGURE 6.

Considering optical excitation at 1.08 eV (1140 nm) and both the excitation and the probe beams travelling towards right (top panel in FIGURE 6), the first graphene layer absorbs about 3% of the excitation light.[34] Since the MoS$_2$ layer is mostly transparent at this excitation photon energy, the second graphene layer receives lesser amount of light than the first graphene layer. Hence, less photocarriers are generated in the second graphene layer. This difference results into a photocarrier concentration gradient across the heterostructure. Of course, the strength of the photocarrier concentration gradient would depend on the excitation photon energy and the fluence. Accordingly, an electric field develops across the GrMoS$_2$Gr heterostructure. It lifts the degeneracy around the K-points and gives rise to a small direct bandgap [23].

Energetic electrons are generated by the excitation pulse (photon energy > 0.5 eV) in the graphene at sufficiently high energy states. Due to the photocarrier concentration gradient, while relaxing, these energetic electrons are dragged towards MoS$_2$ layer during the excitation pulse itself. Taking, the out of plane component of the electronic velocity to be $10^5$ m/s, interlayer separation between graphene and MoS$_2$ to be 0.6 nm, the photocarriers would take a time of ~6 fs, i.e., mainly during the excitation pulse duration itself, to move across the Gr/MoS$_2$ interface. Once in the MoS$_2$ and relaxed to its conduction band minimum, the energetic electrons have to go back towards the graphene layer to be recombined with the corresponding photogenerated holes. Due to the smaller DOS in graphene, the photoelectrons transferred to MoS$_2$ layer would keep piling up there until states are available for their return at lower energies in graphene. The later occurs within the time-scale of electron-hole recombination in graphene. Therefore, the additional pathway in the carrier relaxation trajectories results into slowing down of the overall relaxation dynamics in the GrMoS$_2$Gr through the occurrence of MoS$_2$-like relaxation component ($A_2,\tau_2$) in FIGURES 3, 4 and 5. Clearly, at excitation photon energies below 0.5 eV (energy difference between the bottom of the conduction band of MoS$_2$ and the K-point of graphene), the energetic photoelectrons in the graphene are not created in states sufficiently high in energy so that they can be dragged towards MoS$_2$ due to concentration gradient field. Therefore, for excitation photon energies close and below 0.5 eV (2700 nm), the relaxation of all the photocarriers occurs within the graphene layers only resulting into a transient THz response of the GrMoS$_2$Gr resembling that of the graphene sample.

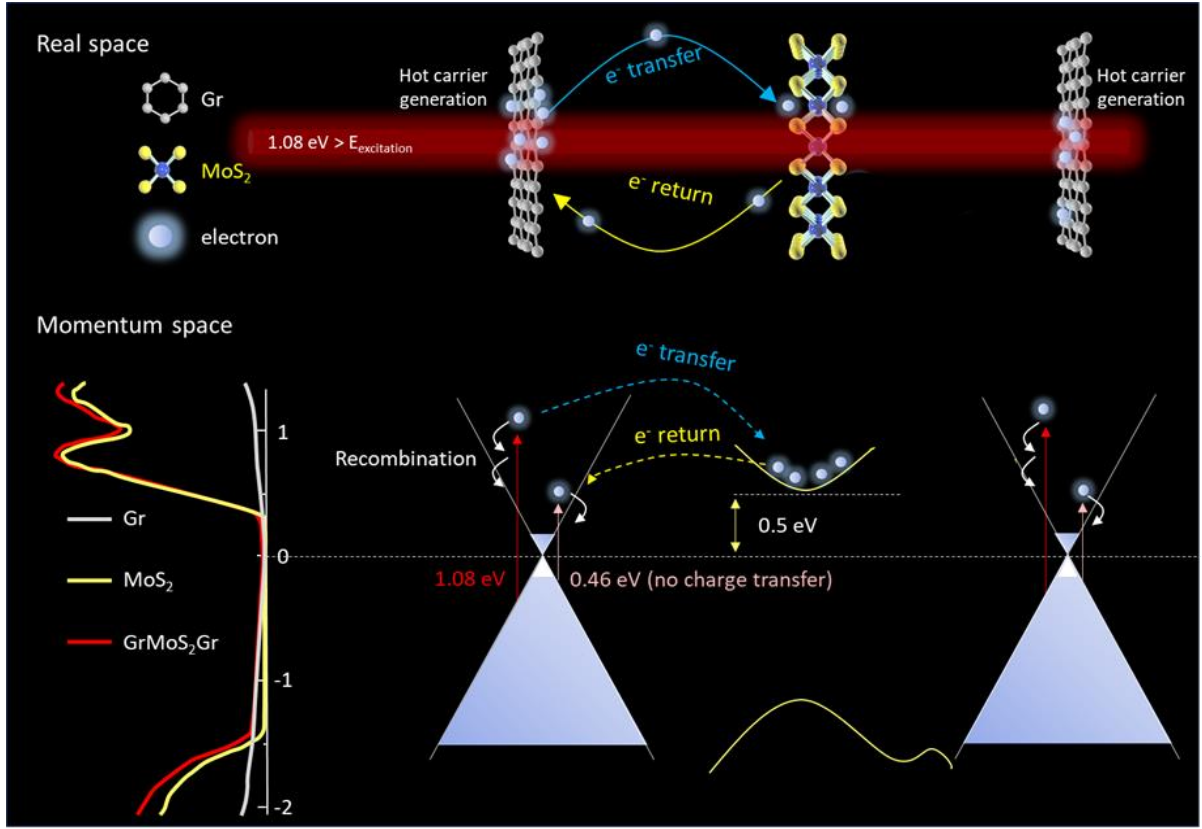

**FIGURE 6** Real space and momentum space representations of the optically excited carrier movement between the graphene and MoS$_2$ layers accounting for the transient THz response of the GrMoS$_2$Gr heterostructure for an optical excitation at 1.08 eV (1140 nm), which creates photocarriers only in the graphene layers of the heterostructure.

Note that for photoexcitation with >0.5 eV photons, the energetic photocarriers move, only partially, towards the MoS$_2$ layer, and a fraction of them still relaxes within the graphene layer. The ratio between the magnitudes of the Gr- and MoS$_2$-like components in the THz response of GrMoS$_2$Gr provides directly the fraction of the photocarriers migrating from the graphene to the MoS$_2$ layer of the heterostructure due to photocarrier concentration gradient in the GrMoS$_2$Gr heterostructure. From FIGURES 3 and 4, we find that at 1.55 eV excitation, about 35% of the photoexcited carriers relax within the graphene layer while about two third of the total, instantly migrate to the MoS$_2$ layer. At optical excitation of ~0.46 eV, the fraction of migrating photocarriers abruptly decreases below 10% (FIGURE 5).

### 4. Conclusions

To summarize, we have presented optically induced transient THz reflectivity and transmission changes and their recovery in high quality graphene, MoS$_2$ single layer and graphene-MoS$_2$-graphene heterostructure. From the GrMoS$_2$Gr heterostructure, an enhanced transient THz response is produced even for excitation photon energies deep in the band gap region of the MoS$_2$ single layer. The excitation photon energy dependent measurements clearly suggest that, only below 0.45 eV, the MoS$_2$-like contribution vanishes from the ultrafast THz response of the heterostructure. This is a clear case of an all-optical control of THz conductivity in MoS$_2$ monolayer even at those optical frequencies for which MoS$_2$ layer is completely transparent, by encapsulating it with graphene from either sides. This effect may have many implications in fast electronic devices at GHz to THz frequencies, such as in FETs, where, without creating electrons within the MoS$_2$ layer by the application of an electrical bias, the external light can help in modulating the conductivity of the MoS$_2$ layer through top and bottom graphene layers. This phenomenon seems to be quite general and should occur in heterostructures for a large variety of TMD layers and other semiconducting layers sandwiched in between two graphene or metallic layers, thereby, opening up new avenues for generation of interesting material heterostructures to enhance certain device performances in the optoelectronics and THz photonics.

### 5. Materials and Experiments

*Sample preparation:* Large area (~15mm × 15mm) MoS$_2$ monolayer was grown on SiO$_2$/Si substrate by two-zone furnace atmospheric pressure CVD. Ammonium heptamolybdate ((NH$_4$)$_6$Mo$_7$O$_{24}$, Sigma-Aldrich, 431346) and sodium hydroxide (NaOH, Sigma-Aldrich, S8045) were dissolved in deionized water as a Mo precursor and promoter, respectively. The precursor solution was coated onto the substrate by spin-coating at 3000 r.p.m. for 1 min. The details are described in our previous work.[35] Then, the precursor-coated substrate and 0.2 g of sulfur (Sigma, 344621) were separately introduced in to a two-zone furnace. The sulfur zone was heated up to 210 °C at a rate of 50 °C min$^{-1}$, while the substrate zone was set to 800 °C with a ramping rate of 80 °C min$^{-1}$. For the entire process, $N_2$ gas (500 sccm) was injected as a carrier gas. A Cu foil (Nilaco, CU-113173, 99.9% with 100 μm thickness) was annealed in $H_2$ and Ar atmosphere at 1050 °C for 2 hours to increase its crystallinity and remove any contamination. 5 sccm of diluted $CH_4$ gas (0.1% diluted in Ar) was then injected into the CVD chamber for 15 min to synthesize graphene on the Cu foil. After the growth, the CVD chamber was naturally cooled to room temperature. For transferring the $MoS_2$ and graphene to quartz substrates, poly(methyl methacrylate) (PMMA C4, MicroChem) assisted wet-etching method was adopted. PMMA was coated onto the sample ($MoS_2$ on the $SiO_2$/Si, and Gr on Cu foil) following which the PMMA-supported samples were immersed into respective etchants (diluted hydrofluoric acid for $MoS_2$ and Cu-etchant for Gr) to detach the samples from the as-grown substrates. The PMMA-supported samples were transferred to quartz substrates. Finally, PMMA was removed by dipping the PMMA-supported samples on quartz substrates into acetone for 5 min. A tri-layer graphene-$MoS_2$-graphene (Gr$MoS_2$Gr) heterostructure was fabricated for our current studies. Initially a monolayer $MoS_2$ was transferred onto the graphene sample. Prior to the stacking of $MoS_2$ and graphene layers, the graphene sample was annealed at 300 °C in a vacuum chamber at a pressure of ∼5 × 10$^{-5}$ Torr for 2 hours to avoid any contamination between the interface. The same procedure was followed for stacking the top graphene layer on the $MoS_2$/graphene sample supported on quartz substrate. Thus obtained large area layers are not continuous entirely, but cover about 80% area on the surface on the substrate.[2] High quality and crystallinity of our layered samples and their vertical stacks were confirmed by various characterization tools.[35] Characterization by Raman and optical absorption spectroscopy are given in the supporting information.

*THz time-domain spectroscopy for broadband static THz conductivity:* The schematic of the experimental setup is shown in FIGURE S3. THz pulses of sub-200 fs duration are generated in dual color air plasma and detected in a <110> cut GaP crystal by electro-optic sampling technique in the usual manner[2]. For driving the THz setup, femtosecond pulses with time-duration of ∼50 fs are taken from a Ti:Sapphire based regenerative amplifier operating at 1 kHz repetition rate and central wavelength (photon energy) 800 nm (∼1.55 eV). A strong part (∼1 W average power) of the laser output is used for THz generation. A biconvex lens of focal length 250 mm is used to focus the fundamental and second harmonic from a beta barium borate (BBO crystal, thickness 100 μm) into the air. Highly resistive silicon wafer was used to block the unused optical beam and let through the THz beam in the main experimental setup until the GaP crystal. A set of four off-axis parabolic mirrors with focal length of 150 mm were used to route the THz beam through a sample, placed at the center between the inner two parabolic mirrors and finally to the 200 μm thick GaP crystal for detection of the THz pulses. A synchronous gating optical beam (average power ∼5 mW) taken from the laser and routed through a linear time-delay stage (controllable time-delay) is made to collinearly fall onto the GaP crystal. The gating pulse samples the THz beam in the GaP crystal through an imbalance in intensity of its two orthogonal polarization components detected in a lock-in amplifier with the help of a quarter wave plate, a Wollaston prism and two photodiodes. The complete THz setup was enclosed in a box under continuous purging of dry air to minimize THz losses in between from the generation point in the air-plasma to the detection point at the GaP crystal.

*Optical pump-THz probe experiments:* The same setup as described above and shown in FIGURE S3, was used for optical pump-THz probe time-resolved spectroscopy by adding a pump pulse travelling through another linear translational stage. For optical excitation of the materials studied in our paper, either 800 nm pulses taken directly from the laser or its second harmonic at wavelength (energy) 400 nm (∼3.10 eV) were used. For optical excitation at wavelengths other than the above, we have used an optical parametric amplifier providing ∼100 fs pulses in a variable wavelength range. The time-delay between the optical pump and a fixed position corresponding to the maximum of the THz pulse, either reflected from or transmitted through the sample under study, was varied to obtain the time-resolved optically induced THz response from the samples in both the THz transmission and reflection modes. In all of the experimental results discussed in our paper, the optical excitation beam diameter on the sample was ∼3 mm while the THz probe beam size was slightly larger. The pump-fluence was varied using continuously variable neutral density filters.

**Supporting Information**

Supporting Information is available from the Wiley Online Library or from the author.

**Acknowledgements**

SK acknowledges SERB, Department of Science and Technology, Govt. of India for financial support through projects ECR/2016/000022 and CRG/2020/000892. IIT Delhi is acknowledged for supporting Femtosecond Spectroscopy and Nonlinear Photonics Laboratory. EF acknowledges supports of the Conseil Regional Nouvelle Aquitaine and FEDER for funding the equipment of the COLA platform. YHL acknowledges the financial support by the Institute for Basic Science (IBS-R011-D1).

**References**

[1] E. Y. Ma, B. Guzelturk, G. Li, L. Cao, Z.-X. Shen, A. M. Lindenberg, T. F. Heinz, *Science Advances* **2019**, *5*, eaau0073.

[2] S. Kumar, A. Singh, S. Kumar, A. Nivedan, M. Tondusson, J. Degert, J. Oberlé, S. J. Yun, Y. H. Lee, E. Freysz, *Opt. Express* **2021**, *29*, 4181.

[3] G. Froehlicher, E. Lorchat, S. Berciaud, *Phys. Rev. X* **2018**, *8*, 011007.

[4] W. Choi, I. Lahiri, R. Seelaboyina, Y. S. Kang, *Critical Reviews in Solid State and Materials Sciences* **2010**, *35*, 52.

[5] S. Tongay, W. Fan, J. Kang, J. Park, U. Koldemir, J. Suh, D. S. Narang, K. Liu, J. Ji, J. Li, R. Sinclair, J. Wu, *Nano Lett.* **2014**, *14*, 3185.

[6] J. Plutnar, M. Pumera, Z. Sofer, *J. Mater. Chem. C* **2018**, *6*, 6082.

[7] L. Yu, Y.-H. Lee, X. Ling, E. J. G. Santos, Y. C. Shin, Y. Lin, M. Dubey, E. Kaxiras, J. Kong, H. Wang, T. Palacios, *Nano Lett.* **2014**, *14*, 3055.

[8] G. W. Shim, K. Yoo, S.-B. Seo, J. Shin, D. Y. Jung, I.-S. Kang, C. W. Ahn, B. J. Cho, S.-Y. Choi, *ACS Nano* **2014**, *8*, 6655.

[9] Y. An, Y. Hou, S. Gong, R. Wu, C. Zhao, T. Wang, Z. Jiao, H. Wang, W. Liu, *Phys. Rev. B* **2020**, *101*, 075416.

[10] Y. An, Y. Hou, K. Wang, S. Gong, C. Ma, C. Zhao, T. Wang, Z. Jiao, H. Wang, R. Wu, *Advanced Functional Materials* **2020**, *30*, 2002939.

[11] J. Yoon, W. Park, G.-Y. Bae, Y. Kim, H. S. Jang, Y. Hyun, S. K. Lim, Y. H. Kahng, W.-K. Hong, B. H. Lee, H. C. Ko, *Small* **2013**, n/a.

[12] M. Zhao, Y. Ye, Y. Han, Y. Xia, H. Zhu, S. Wang, Y. Wang, D. A. Muller, X. Zhang, *Nature Nanotech* **2016**, *11*, 954.

[13] C.-J. Shih, Q. H. Wang, Y. Son, Z. Jin, D. Blankschtein, M. S. Strano, *ACS Nano* **2014**, *8*, 5790.

[14] Yuchen Du, Lingming Yang, Jingyun Zhang, Han Liu, K. Majumdar, P. D. Kirsch, P. D. Ye, *IEEE Electron Device Lett.* **2014**, *35*, 599.

[15] W. J. Yu, Y. Liu, H. Zhou, A. Yin, Z. Li, Y. Huang, X. Duan, *Nature Nanotech* **2013**, *8*, 952.

[16] H. Wang, H. Feng, J. Li, *Small* **2014**, *10*, 2165.

[17] K. Roy, M. Padmanabhan, S. Goswami, T. P. Sai, G. Ramalingam, S. Raghavan, A. Ghosh, *Nature Nanotech* **2013**, *8*, 826.

[18] X. Sun, B. Zhang, Y. Li, X. Luo, G. Li, Y. Chen, C. Zhang, J. He, *ACS Nano* **2018**, *12*, 11376.

[19] F. H. L. Koppens, T. Mueller, Ph. Avouris, A. C. Ferrari, M. S. Vitiello, M. Polini, *Nature Nanotech* **2014**, *9*, 780.

[20] Y. Ma, Y. Dai, M. Guo, C. Niu, B. Huang, *Nanoscale* **2011**, *3*, 3883.

[21] B. Qiu, X. Zhao, G. Hu, G. Yue, J. Ren, X. Yuan, *Nanomaterials* **2018**, *8*, 962.

[22] C. Xia, W. Xiong, W. Xiao, J. Du, L. Fang, J. Li, Y. Jia, *Phys. Rev. Applied* **2018**, *10*, 024028.

[23] L. Shao, G. Chen, H. Ye, Y. Wu, H. Niu, Y. Zhu, *EPL* **2014**, *106*, 47003.

[24] B. Sachs, L. Britnell, T. O. Wehling, A. Eckmann, R. Jalil, B. D. Belle, A. I. Lichtenstein, M. I. Katsnelson, K. S. Novoselov, *Appl. Phys. Lett.* **2013**, *103*, 251607.

[25] D. Kozawa, R. Kumar, A. Carvalho, K. Kumar Amara, W. Zhao, S. Wang, M. Toh, R. M. Ribeiro, A. H. Castro Neto, K. Matsuda, G. Eda, *Nat Commun* **2014**, *5*, 4543.

[26] L. Wang, Z. Wang, H.-Y. Wang, G. Grinblat, Y.-L. Huang, D. Wang, X.-H. Ye, X.-B. Li, Q. Bao, A.-S. Wee, S. A. Maier, Q.-D. Chen, M.-L. Zhong, C.-W. Qiu, H.-B. Sun, *Nat Commun* **2017**, *8*, 13906.

[27] R. Zan, Q. M. Ramasse, R. Jalil, T. Georgiou, U. Bangert, K. S. Novoselov, *ACS Nano* **2013**, *7*, 10167.

[28] Z. Sun, H. Chu, Y. Li, S. Zhao, G. Li, D. Li, *Materials & Design* **2019**, *183*, 108129.

[29] L. Britnell, R. V. Gorbachev, R. Jalil, B. D. Belle, F. Schedin, A. Mishchenko, T. Georgiou, M. I. Katsnelson, L. Eaves, S. V. Morozov, N. M. R. Peres, J. Leist, A. K. Geim, K. S. Novoselov, L. A. Ponomarenko, *Science* **2012**, *335*, 947.

[30] K. J. Tielrooij, J. C. W. Song, S. A. Jensen, A. Centeno, A. Pesquera, A. Zurutuza Elorza, M. Bonn, L. S. Levitov, F. H. L. Koppens, *Nature Physics* **2013**, *9*, 248.

[31] C. J. Docherty, P. Parkinson, H. J. Joyce, M.-H. Chiu, C.-H. Chen, M.-Y. Lee, L.-J. Li, L. M. Herz, M. B. Johnston, *ACS Nano* **2014**, *8*, 11147.

[32] C. H. Lui, A. J. Frenzel, D. V. Pilon, Y.-H. Lee, X. Ling, G. M. Akselrod, J. Kong, N. Gedik, *Phys. Rev. Lett.* **2014**, *113*, 166801.

[33] S. S. Baik, S. Im, H. J. Choi, *Sci Rep* **2017**, *7*, 45546.

[34] V. G. Kravets, A. N. Grigorenko, R. R. Nair, P. Blake, S. Anissimova, K. S. Novoselov, A. K. Geim, *Phys. Rev. B* **2010**, *81*, 155413.

[35] S. J. Yun, G. H. Han, H. Kim, D. L. Duong, B. G. Shin, J. Zhao, Q. A. Vu, J. Lee, S. M. Lee, Y. H. Lee, *Nature Communications* **2017**, *8*, 1.

# Supporting Information

# Sub-bandgap activated charges transfer in a graphene-MoS$_2$-graphene heterostructure

Sunil Kumar,* Arvind Singh, Anand Nivedan, Sandeep Kumar, Seok Joon Yun, Young Hee Lee, Marc Tondusson, Jérôme Degert, Jean Oberle, and Eric Freysz*

[1]*Femtosecond Spectroscopy and Nonlinear Photonics Laboratory,
Department of Physics, Indian Institute of Technology Delhi, New Delhi 110016, India*
[2]*Department of Energy Science, Sungkyunkwan University (SKKU), Suwon 16419, Republic of Korea*
[3]*Univ. Bordeaux, CNRS, LOMA UMR 5798, 33405 Talence, France*
*kumarsunil@physics.iitd.ac.in; *eric.freysz@u-bordeaux.fr

Content:

S1: *Raman spectroscopic characterization of the layers.*

S2: *Characterization of the layers by optical absorption spectroscopy.*

S3: *Experimental setup*

S4: *Broadband static THz sheet conductivity of the thin layers.*

S5: *Fitting of optical pump-induced THz reflectivity and transmission data.*

S6: *Transient THz response of optically pumped graphene.*

S7: *Transient THz response of MoS$_2$ monolayer at 3.1 eV optical pumping.*

S8: *Transient THz transmissivity of GrMo$_2$Gr heterostructure at 1.55 eV optical excitation.*

S9. *Transient THz reflectivity of GrMo$_2$Gr heterostructure at 3.1 eV optical excitation*

## S1. Raman spectroscopic characterization of the layers

Raman spectroscopy was used to confirm the high quality of our samples. A Renishaw spectrometer at 532 nm laser excitation was used to record the Raman spectra. The results for the graphene, $MoS_2$ monolayer and $GrMoS_2Gr$ heterostructure are described in FIGURE S1(a) in the entire experimental range from 280 cm$^{-1}$ to 3000 cm$^{-1}$ revealing all the characteristic phonon lines of the graphene and $MoS_2$ monolayer as indicated. All these characteristic phonon frequencies of graphene and $MoS_2$ monolayer are consistent with the literature. The 2D-mode and G-mode of graphene are observed at frequencies ~1587 cm$^{-1}$ and 2680 cm$^{-1}$, respectively. A Raman peak at frequency between 1300 and 1350 cm$^{-1}$ corresponding to the defects associated D-band in finite size graphene is hardly seen in the Raman spectra of either graphene or $GrMoS_2Gr$ (inset in FIGURE S1(a)). Therefore, an intensity ratio, $I_D/I_G \ll 1$ evidently shows an extremely high-quality graphene used in our experiments.

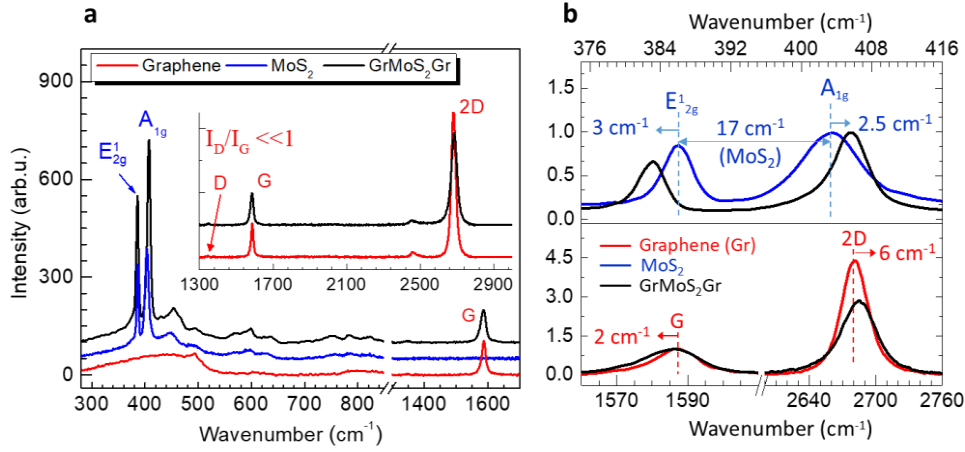

**FIGURE S1** Characterization of graphene, $MoS_2$ monolayer and the graphene-$MoS_2$-graphene ($GrMoS_2Gr$) heterostructure using Raman spectroscopy. **a)** Raman spectra of the three samples revealing all the characteristic phonon lines of the graphene and the $MoS_2$ monolayer. **b)** Details and analysis of G- and 2D peaks of graphene, and $E_{2g}^1$ and $A_{1g}$ modes of $MoS_2$ monolayer. The shifts in the frequencies of these modes in the $GrMoS_2Gr$ heterostructure have been marked. Graphene: G-mode at ~1587 cm$^{-1}$, 2D-mode at ~2680 cm$^{-1}$; $MoS_2$: $E_{2g}^1$-mode at ~386 cm$^{-1}$ and $A_{1g}$ mode at ~403 cm$^{-1}$.

The well known $E_{2g}^1$ and $A_{1g}$ symmetric phonons of $MoS_2$ monolayer are clearly visible at frequencies of ~386 cm$^{-1}$ and 403 cm$^{-1}$, respectively, and match very well with the literature.[33] It is known from the literature[33] that with the increase in the number of layers in the $MoS_2$ samples, the difference in the frequencies of these two phonon modes keeps increasing, while the minimum of ~17 cm$^{-1}$ is found for a monolayer $MoS_2$ sample. This is in excellent agreement with our observation in FIGURE S1(b) for the $MoS_2$ monolayer used in the current study.

Due to the van der Walls (vdW) interaction between different layers in the heterostructure, shifts in the frequencies of $E_{2g}^1$ and $A_{1g}$ Raman modes of the $MoS_2$ monolayer and the G- and 2D-modes of graphene are observed, which have been analyzed in FIGURE S1(b). We can see that the Raman peaks due to the $E_{2g}^1$ and $A_{1g}$ modes of $MoS_2$ monolayer exhibit counter frequency shifts in the $GrMoS_2Gr$ heterostructure. The symmetric in-plane $E_{2g}^1$ mode shows a red shift by ~3 cm$^{-1}$, while the out-of-plane $A_{1g}$ mode is blue shifted by about 2.5 cm$^{-1}$ in the $GrMoS_2Gr$ heterostructure. A slight softening of the in-plane G-mode of graphene by ~2 cm$^{-1}$ in the $GrMoS_2Gr$ heterostructure is along the expected lines and behaves in the same manner as the symmetric in-plane $E_{2g}^1$ mode of the $MoS_2$ monolayer. The 2D-mode of graphene blue shifts by a larger amount of ~6 cm$^{-1}$. These observed shifts in the Raman frequencies of $E_{2g}^1$ and $A_{1g}$ modes due to the capping layers (graphene in our case) quite well resemble the observations previously reported.[34–37] Therefore, the overall frequency shifts of the 2D-mode and G-mode of graphene as well as the $E_{2g}^1$ and $A_{1g}$ modes of $MoS_2$ monolayer can be accounted for by the compression and stretching of the graphene and $MoS_2$ monolayer within the $GrMoS_2Gr$ heterostructure.[34] The broadening of the G-peak of graphene in $GrMoS_2Gr$ may be due to the photoelectrons generated by the above bandgap laser excitation of $MoS_2$ monolayer at ~532 nm in the Raman measurements which subsequently get injected into graphene.[38]

## S2. Characterization of the layers by optical absorption spectroscopy

UV-visible absorption spectroscopy also was used to characterize the $MoS_2$ monolayer, graphene and the graphene-$MoS_2$-graphene heterostructure, $GrMoS_2Gr$. From FIGURE 2, the usual π-π* transition band[41] of graphene is observed at ~4.65 eV. The strong absorption features at energies ~1.89 eV, 2.04 eV and 2.87 eV due to A-, B- and C-excitons, respectively, in the $MoS_2$ monolayer, can also be clearly seen in FIGURE S2. The peak positions of the Excitons in $MoS_2$ monolayer are again in excellent agreement with the literature.[39],[40] The energy separation between the A- and B-excitons of the MoS$_2$ monolayer remains nearly the same, however, both redshift by an amount ~0.03 eV in the GrMoS$_2$Gr heterostructure (see lower panel of FIGURE S2). On the other hand, no apparent shift in the C-exciton absorption peak is observed. The weak redshift in the A- and B-exciton peak energies in the GrMoS$_2$Gr heterostructure can be attributed to the interlayer vdW interaction, similar to one in the multilayered and bulk MoS$_2$ samples.[39] The least value of the energy separation of ~0.15 eV for the two peaks is known for a monolayer sample[39]. The same value can be observed from FIGURE S2 to evidently show the monolayer nature of our MoS$_2$ sample.

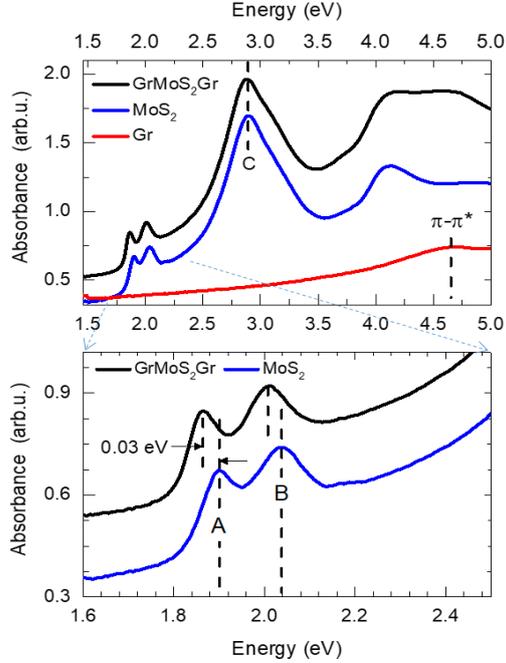

**FIGURE S2** Characterization of the MoS$_2$ monolayer, graphene and graphene-MoS$_2$-graphene (GrMoS$_2$Gr) heterostructure by UV-Visible absorption spectroscopy.

The strong absorption features due to A-, B- and C-excitons in the MoS$_2$ monolayer remain intact in the GrMoS$_2$Gr heterostructure, which suggests that the preparation of the heterostructure has not affected the structural properties of the MoS$_2$ layer.[44] Therefore, the contribution from any additional defect states in the sub-band gap region of MoS$_2$ in the heterostructure cannot be invoked while comparing our transient THz response from MoS$_2$ monolayer and the heterostructure. We have also noticed from the static THz conductivity of the MoS$_2$ monolayer being zero that there are no conducting defect states present. It may be noted that light emitting properties of the defect states have very large time-scale, typically in nanoseconds.

**S3. Experimental setup for THz-TDS and optical pump-THz probe spectroscopy**

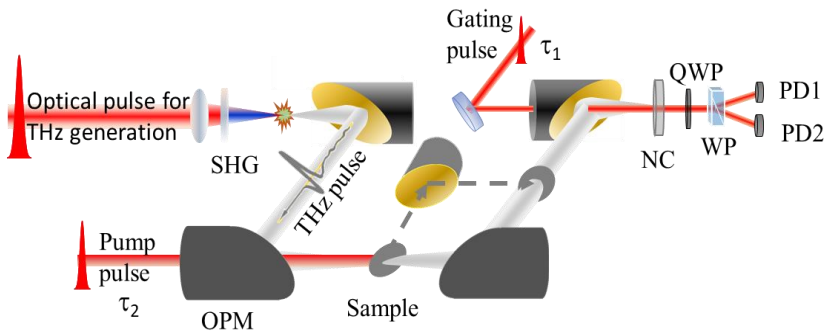

**FIGURE S3** Experimental setup for THz time-domain spectroscopy (THz-TDS) and optical pump-THz probe time-resolved spectroscopy. The THz pulses are generated by dual color air-plasma and detected in a thin GaP crystal by electro-optic sampling.

**S4. Broadband static THz sheet conductivity measurements**

The THz traces in air and transmitted through the quartz substrate are shown in FIGURE S3(a), and the corresponding spectra shown in S3(b). These can be analyzed[45] to determine the complex transmission spectra and hence the frequency-dependent spectra for the index of refraction (n) and the extinction coefficient (κ) of quartz, as shown in FIGURE S3(c). The time-domain traces of the THz radiation transmitted through our graphene, MoS$_2$ monolayer and the GrMoS$_2$Gr heterostructure, all supported on quartz substrates, are shown in FIGURE

S4(d) and the corresponding Fourier spectra are given in FIGURE S4(e). For a comparison, the time-domain scan and the spectrum for THz pulses without any sample (air) are also shown in FIGURES S4(d-e), where, it can be seen that almost 50% reduction in the THz signal is primarily due to the underlying quartz substrate. The difference in the THz Fourier spectra of graphene, $MoS_2$ and $GrMoS_2Gr$ cannot be very well noticed from the results in Figure S4(e) due to the extremely weak THz absorption and dispersion introduced by these atomically thin samples. From the time-domain traces around the maximum (FIGURE S4(d) and the inset), it is apparent that the THz absorption by graphene and $GrMoS_2Gr$ is substantially larger than that by the $MoS_2$ monolayer. In fact, as it becomes clear from the conductivity results also, discussed in the main paper, the THz absorption by the $MoS_2$ monolayer is negligibly small in the entire experimental THz frequency window of our experiments. We would also like to point out that the THz time-domain traces for the layers supported on quartz substrate show the maximum signal appearing at different times for all of them (see FIGURE S4(d) and the zoomed-in view in the inset). This is due to the fact that the thickness of the quartz substrate in each case was slightly different. The independently measured thickness of the substrate for each case was found to be exactly the same as the one calculated from the temporal shift and the mean value of the refractive index (FIGURE S4(c)) of quartz.

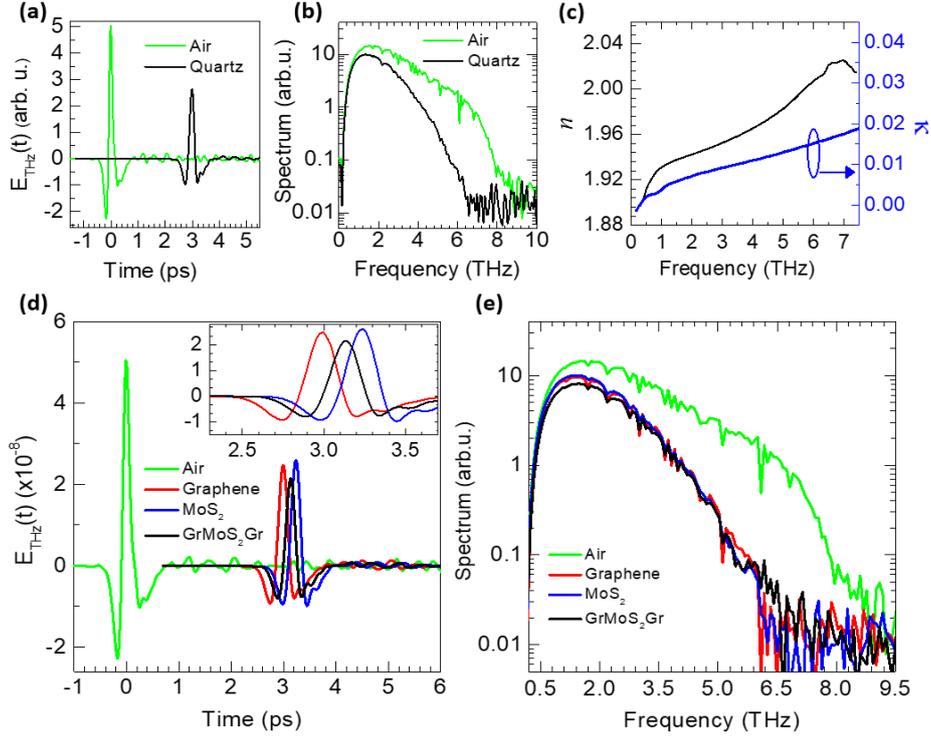

**FIGURE S4** Experimental time-domain THz traces and the Fourier domain spectra from the samples and without any sample in air. **a-b)** THz traces for the quartz plate that was used as substrate for our layer samples and for the air as reference. **c)** Index of refraction, $n$ and the extinction coefficient, $\kappa$ obtained for the quartz plate following thin film analysis as described in the text. **d-e)** THz time-domain scans and the corresponding Fourier transforms obtained on difference samples; same are shown for air as reference. The inset in **(d)** is the zoomed-in view around the field maximum to show the temporal shift between the scans for different samples due to the slight difference in the thickness of the quartz substrate used in them.

The complex transmission spectra were also obtained for the samples (atomically thin films on the quartz substrate) from the respective experimental data. By applying the standard thin–film approximation, we have extracted the frequency-dependent complex sheet conductivity, σ(ω), of our thin layers using the following relation[42]

$$\sigma(\omega) = c\varepsilon_0[1 + \tilde{n}_{sub}(\omega)]\left(\frac{1}{T_{film}(\omega)} - 1\right) \tag{S.1}$$

where, $T_{film}(\omega)$ is the complex transmission coefficient of the film (material layer) alone, $\varepsilon_0$ is free space permittivity, and c is speed of light in vacuum. $T_{film}(\omega)$ is calculated from the experimentally known transmission coefficient of the sample (film + substrate) and the substrate alone, i.e., $T_{film}(\omega) = T_{sample}(\omega)/T_{sub}(\omega)$. Substituting the values of $\tilde{n}_{sub}(\omega)$ and $T_{film}(\omega)$ into Equation (S.1), the complex sheet conductivity is calculated, whose real part, Re(σ), and imaginary part, Im(σ), for each sample, have been presented in the main paper. These could be evaluated reliably only up to a maximum frequency of ~5 THz due to the strong absorption in the underlying quartz substrate at higher frequencies. By using another substrate for the films that is transparent in broader THz window, one could extract the THz conductivity in the entire experimental frequency range. We may note that the sheet conductivity of the samples, as given in the main paper, already includes the thickness of the layers and hence its unit is in S or milli-S. By using appropriate values for the thickness of each layer, results in absolute S/m can be obtained.

## S5. Fitting of optical pump-induced THz reflectivity and transmission data

The temporal evolution of the optical pump-induced THz response, i.e., the ΔR and ΔT can be analyzed simply by numerically fitting the experimental data using the following response function:

$$Y(t) = \left( H(t) \times \sum_i A_i e^{-t/\tau_i} \right) \otimes \exp(-t^2/\tau_p^2) \tag{S.2}$$

The first function, H(t) on the right side of the above equation is the standard Heaviside step function to account for material response at only positive times from the instance the optical pump excites the sample. The temporal evolution of the material response is captured mainly by the second function, which is a sum of exponentially decaying functions having respective amplitude, $A_i$ and time-constant, $\tau_i$. Finally, convolution of the product of the above two functions is taken with the Gaussian probe pulse having time-duration $\tau_p = 200$ fs, as described in Equation (S.2) to numerically fit the experimental data in our study using MATLAB.

## S6. Transient THz response of optically pumped graphene

Ultrafast photocarrier dynamics in graphene is well known in the literature. In our paper, the ultrafast THz response of graphene has been investigated using optical excitation at 800 nm (1.55 eV), 400 nm (3.1 eV), 1140 nm (1.08 eV), 1240 nm (0.9 eV) and 2700 nm (0.45 eV). The response from THz reflectivity and transmission changes at optical excitation 3.1 eV and 1.55 eV are given in FIGURE S6. At both photon energies, an initial decrease in the THz reflectivity is observed which recovers completely within 7 ps at all the experimental pump-fluences. The experimental data for consecutive pump-fluences have been shifted vertically for clarity, in the upper panels of FIGURE S6, while the thick continuous lines are representative curves obtained by numerical fits of the data using the procedure described above. It can be seen that only a single exponential function is enough to fit the experimental data completely. The pump-fluence dependence of the corresponding kinetic parameters, i.e., the amplitudes, $A_i$ and the time-constants, $\tau_i$ of the exponentially decaying recovery of the data have been presented in the lower panels of FIGURE S6.

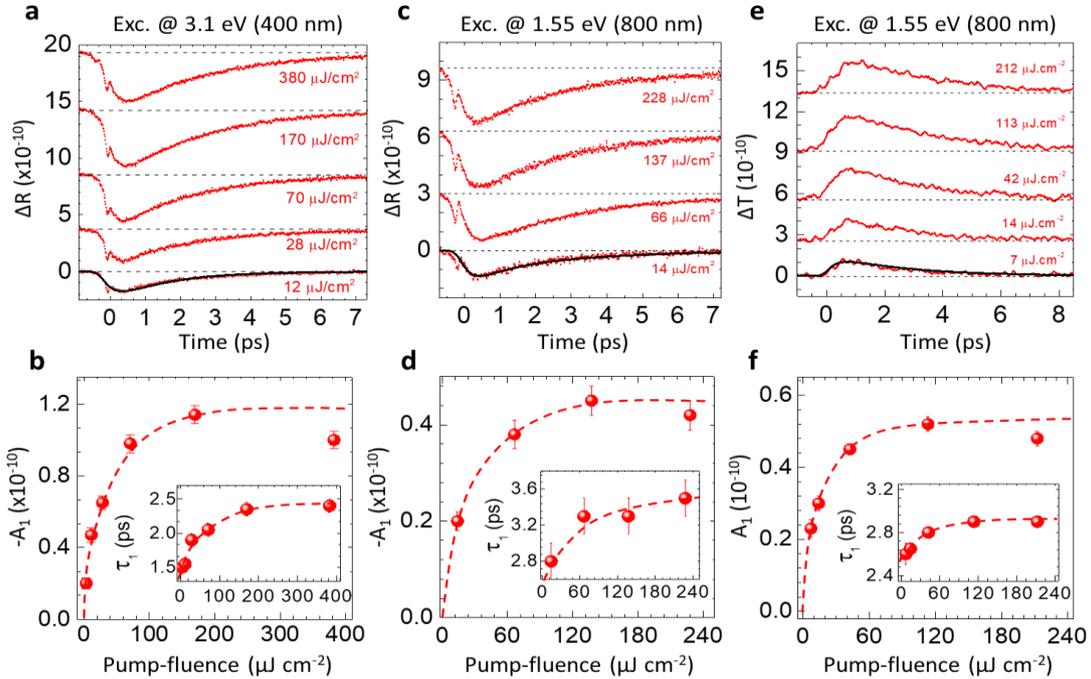

**FIGURE S6** Ultrafast THz response of graphene from through ΔR and ΔT kinetics at optical pump photon energies of 3.1 eV and 1.55 eV. Thick continuous lines in the time-resolved spectra are numerical fits obtained using exponentially decaying function convolved with a Gaussian probe pulse of 200 fs FWHM duration. The thick dashed curves in lower panels represent mean behavior of the fluence dependence of the kinetic parameters. **a-b)** ΔR kinetics at 3.1 eV (400 nm) at various optical pump-fluences and the corresponding amplitude $A_1$ and decay time-constant $\tau_1$ of the relaxation. **c-d)** ΔR kinetics at 1.55 eV (800 nm) at varying optical pump-fluences and the corresponding amplitude $A_1$ and decay time-constant $\tau_1$ of the relaxation. **e-f)** ΔT kinetics at 1.55 eV (800 nm) at various optical pump-fluences and the corresponding amplitude $A_1$ and decay time-constant $\tau_1$ of the relaxation.

Saturation of the ultrafast optical pump-induced THz response at a fluence of ~60 μJ/cm$^2$ is observed from FIGURES S6(b) and S(d) at the pump photon energies of 3.1 eV and 1.55 eV, respectively. The corresponding time-constants are shown in the insets of FIGURES S6(b) and S6(d), respectively. The dashed lines in FIGURES S6(b) and S6(d) have been drawn as guide to the eyes and represent the mean behavior of the kinetic parameters. The relaxation time-constant initially increases with the increasing pump-fluence and then attain a constant value beyond the saturation fluence. From these results, it also appears that at any given pump-fluence, the relaxation is faster at 3.1 eV than at 1.55 eV optical excitation. The relaxation time-constant in graphene is found to be in between 1.5 ps and 3.5 ps at any optical excitation photon energy in our experiments which is quite consistent with the reports in the literature. We may mention that another much faster sub-500 fs time-constant is also known to be present in the hot carrier dynamics in graphene. However, our current experimental data could not resolve it.

The optically induced THz response from the ΔR and ΔT measurements is exactly the same except much higher signal to noise ratio and an opposite sign of the signal in the case of ΔR. This can be compared, for example, from FIGURES S6(c) and S6(e), where results at 1.55 eV excitation for the ΔR and ΔT response have been presented at various pump-fluences and the corresponding amplitude, A and time-constant, τ of the relaxation channel are given in FIGURES S6(d) and S6(f), respectively. Moreover, a small coherent artifact appearing near the initial decay of the ΔR signals can also be noticed from FIGURES S6(a) and S6(c) which, however, does not affect our fitting procedure or the conclusions drawn in our paper and, hence, can be ignored without further discussion. Therefore, we have used ΔR mostly in our main paper to analyze the optically induced THz response from our materials.

**S7. Transient THz response of MoS$_2$ monolayer at 3.1 eV optical pumping**

Optical excitation photon energy of 3.1 eV is much higher than the band gap energy of MoS$_2$ monolayer, hence photocarriers are generated, which modulate the THz response of the monolayer. The same is not true for the below band gap excitation at 1.55 eV and hence no signal can be obtained in the time-resolved measurements at this optical excitation as clearly seen from our experiments. The results for the ultrafast THz ΔR response of MoS$_2$ monolayer at 3.1 eV excitation photon energy and various pump-fluences are presented in FIGURE S7. The results clearly indicate the presence of a two-step relaxation of the photoinduced recovery while the signals do not completely vanish within the experimental window up to 9 ps. The other major difference in these signals from those obtained for graphene at the same excitation photon energy (FIGURE S6) is the opposite sign of the response, i.e., an initial increase of the ΔR signals. By numerical fitting of the data (fits shown by dark continuous lines in FIGURE S7(a)), we obtain the amplitudes and the corresponding time-constants for the two relaxation channels. These kinetic fit parameters are presented in FIGURE S7(b) and S7(c) as a function of the pump-fluence, where dashed curves have also been drawn as guide to the eyes and represent the mean behavior of the fluence-dependence.

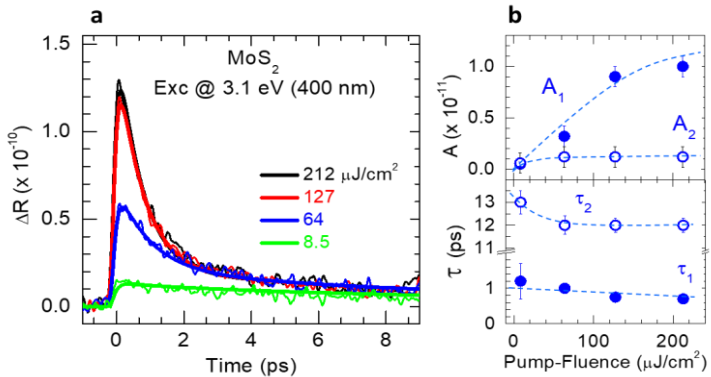

**FIGURE S7** Ultrafast THz response of MoS$_2$ monolayer at excitation photon energy of 3.1 eV (wavelength 400 nm). **a)** Temporal evolution of differential THz reflectivity at various pump-fluences. The thick continuous lines are numerical fits of the data to extract the kinetic parameters of the evolution. **b)** Pump-fluence dependence of the amplitudes, $A_i$ and corresponding time-constants, $τ_i$ of the relaxation components in the THz response. The dashed lines represent the mean behavior of the parameters.

**S8. Transient THz transmission of GrMo$_2$Gr heterostructure at 1.55 eV optical excitation**

Using optical excitation photon energy of 1.55 eV, which is within the sub-bandgap region of the MoS$_2$ monolayer, the transient THz response through ΔT, are presented in FIGURE S8. The time-resolved data along with numerical fits are presented in FIGURE S8(a). The corresponding fit parameters of two relaxation channels are shown in FIGURES S8(b) and S8(c), where dashed lines have been drawn as guide to the eyes to show the mean behavior of the parameters. The data show a negative contribution at shorter times and another positive component at longer times. We may recall that at excitation photon energy of 1.55 eV, a positive ΔT response is characteristic to the graphene. However, the negative component at long times in ΔT resembles that of MoS$_2$ monolayer but at an excitation photon energy of 3.1 eV. The details of the possible origin to this fact have been discussed in the main paper. Therefore, the THz ΔT response of GrMoS$_2$Gr has two contributions in its temporal evolution, one is graphene-like (A$_1$,τ$_1$) and another MoS$_2$-like (-A$_2$,τ$_2$).

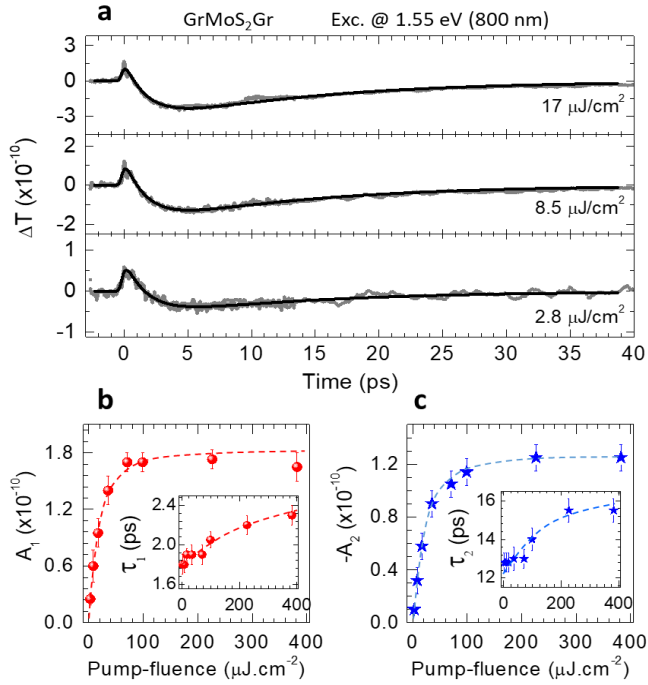

**FIGURE S8** Ultrafast THz response of graphene-MoS$_2$-graphene (GrMoS$_2$Gr) heterostructure through measurement of pump-induced ΔT at optical excitation of 1.55 eV (800 nm). **a)** Time-resolved spectra at various pump-fluences. Thick continuous lines are results from numerical fitting of the data. **b-c)** Pump-fluence dependence of the relaxation parameters: amplitudes, A$_i$ and time-constants τ$_i$. A$_1$ and τ$_1$ correspond to the graphene-like response and A$_2$ and τ$_2$ are parameters of the MoS$_2$-like relaxation component in the overall response. Dashed lines in the lower two panels represent the mean behavior of the fluence-dependence of the relaxation parameters.

## S9. Transient THz reflectivity of GrMo$_2$Gr heterostructure at 3.1 eV optical excitation

The transient THz reflectivity induced by optical pump at 1.55 eV has clearly shown Gr- and MoS2-like components. For optical excitation at 3.1 eV, the raw traces taken at various pump-fluences are presented in FIGURE S9(a). Similar to the previous case, the experimental data is fitted by using a bi-exponentially decaying function. The corresponding amplitudes and time-constants of the two-components are shown in FIGURES S9(b-d), where, the fast negative component is due to graphene and slow positive one is due to MoS$_2$. In the raw traces, though the data beyond 0.8 ps is quite well reproduced by the fitting procedure, but near the zero-delay, there appears to be a contribution from a THz pulse-like response in the ΔR signalS. This additional contribution may arise due to THz emission from the optical pump-induced charge separation at the interface which has been seen earlier in MoS$_2$/MoSe$_2$ heterostructure.[43]

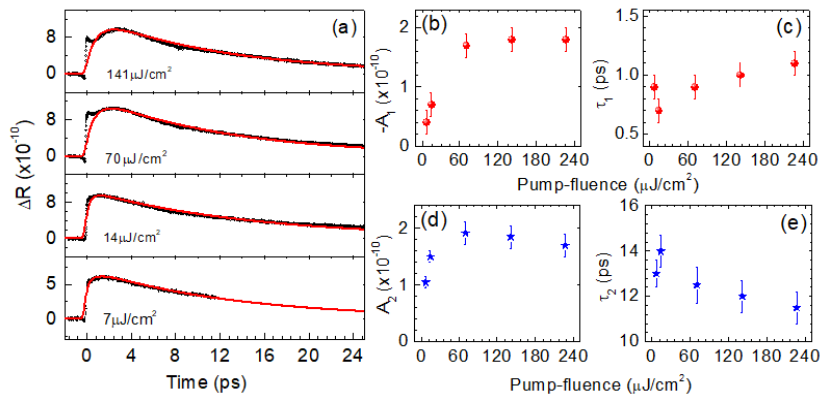

**FIGURE S9 Pump fluence dependence of the u**ltrafast THz response of GrMoS$_2$Gr heterostructure at 3.1 eV (400 nm). **a)** Time-resolved spectra at various pump-fluences. Thick continuous lines are results from numerical fitting of the data at various pump-fluences. **b-e)** Pump-fluence dependence of the amplitudes and time-constants of the fast (negative) and slow (positive) relaxation components.

## References

[33]  M. Buscema, et al., *Nano Res.* **2014**, *7*, 561.
[34]  K.-G. Zhou, F. Withers, Y. Cao, S. Hu, G. Yu, C. Casiraghi, *ACS Nano* **2014**, *8*, 9914.
[35]  L. Wang, et al., *Nanoscale* **2017**, *9*, 10846.
[36]  M. Buscema, et al., *Nano Res.* **2014**, *7*, 561.

[37]   Y. Liu, et al., *Front. Phys.* **2019**, *14*, 13607.
[38]   A. Das, et al., *Nature Nanotech* **2008**, *3*, 210.
[39]   K. P. Dhakal, et al., *Nanoscale* **2014**, *6*, 13028.
[40]   L. Wang, et al., *Nature Communications* **2017**, *8*, 1.
[41]   K. F. Mak, L. Ju, F. Wang, T. F. Heinz, *Solid State Communications* **2012**, *152*, 1341.
[42]   H. J. Joyce, et al., *Semicond. Sci. Technol.* **2016**, *31*, 103003.
[43]   E. Y. Ma et al., *Science Advances* **2019**, 5, eaau0073.
[44]   J. Liu,   et al., Journal of Materials Chemistry C **2017**, 5, 11239-11245.
[45]   S. Kumar, et al., *Opt. Express* **2021**, *29*, 4181.